\pdfoutput=1

\documentclass[aps,prb,superscriptaddress,amsfonts,amsmath,amssymb,showpacs,floatfix,reprint]{revtex4-2}

\usepackage[utf8]{inputenc}
\usepackage{times,bbm}
\usepackage{amsfonts}
\usepackage[dvipsnames]{xcolor}
\usepackage{mathtools}
\usepackage{physics}

\usepackage{algorithm}
\usepackage{algorithmic}

\usepackage{times}
\usepackage[normalem]{ulem}
\usepackage{graphicx}
\usepackage{float}
\usepackage{csquotes}

\usepackage{url}
\usepackage{bm}
\usepackage{graphicx}
\usepackage{amsmath}
\usepackage{amstext}
\usepackage{amssymb}
\usepackage{amsfonts}
\usepackage{amsbsy}
\usepackage{verbatim}
\usepackage{color}
\usepackage{tikz}
\usepackage{gensymb}
\usepackage{textcomp}
\usepackage{enumitem}
\usepackage{cancel}

\usepackage[dvipsnames]{xcolor}
\definecolor{newcolor}{RGB}{0,0,0}
\definecolor{darkgreen}{RGB}{0,150,0}

\usepackage{hyperref}
\hypersetup{
	breaklinks,
	colorlinks,
	linkcolor=Mahogany,
	citecolor=NavyBlue,
	urlcolor=NavyBlue,
	pdftitle={},
	pdfauthor={}
}

\usepackage{dsfont}

\newcommand{\imbalance}{\mathcal{I}}
\newcommand{\HXXZ}{H_{\text{XXZ}}}
\newcommand{\HsMBL}{H_{\text{sMBL}}}
\newcommand{\Hdoping}{H_{\text{doping}}}

\definecolor{jens}{rgb}{0,0,0}
\definecolor{cadmiumgreen}{rgb}{0.0, 0.42, 0.24}

\definecolor{augustine}{rgb}{0,0.5,1}

\begin{document}

\title{A route towards engineering many-body localization in real materials}  

\author{A. Nietner}
\affiliation{Helmholtz Zentrum  Berlin, 14109 Berlin, Germany}

\affiliation{Dahlem Center for Complex Quantum Systems, Freie Universit{\"a}t Berlin, 14195 Berlin, Germany}

\author{A. Kshetrimayum}
\affiliation{Helmholtz Zentrum Berlin, 14109 Berlin, Germany}

\affiliation{Dahlem Center for Complex Quantum Systems, Freie Universit{\"a}t Berlin, 14195 Berlin, Germany}

\author{J. Eisert}
\affiliation{Dahlem Center for Complex Quantum Systems, Freie Universit{\"a}t Berlin, 14195 Berlin, Germany}
\affiliation{Helmholtz Zentrum Berlin, 14109 Berlin, Germany}

\author{B. Lake}
\affiliation{Helmholtz Zentrum Berlin, 14109 Berlin, Germany}

\date{\today}

\begin{abstract}
The interplay of interactions and disorder in a quantum many body system may lead to the elusive phenomenon of many body localization (MBL). It has been observed under precisely controlled conditions in synthetic quantum many-body systems, but to detect it in actual quantum materials seems challenging. In this work, we present a path to synthesize real materials that show signatures of many body localization by mixing different species of materials in the laboratory. To provide evidence for the functioning of our approach, we perform a detailed tensor-network based numerical analysis to study the effects of various doping ratios of the constituting materials. Moreover, in order to provide guidance to experiments, we investigate different choices of actual candidate materials. To address the challenge of how to achieve stability under heating, we study the effect of the electron-phonon coupling, focusing on effectively one dimensional materials embedded in one, two and three dimensional lattices. We analyze how this coupling affects the MBL and provide an intuitive microscopic description of the interplay between the electronic degrees of freedom and the lattice vibrations.
Our work provides a guideline for the necessary conditions on the properties of the ingredient materials and, as such,
serves as a road map to experimentally synthesizing real
quantum materials exhibiting signatures of MBL.
\end{abstract}
\maketitle

\emph{Many-body localization} (MBL) is an intriguing phenomenon characterized by the absence of thermalization~\cite{MBLRMP} and conservation of information about initial conditions beyond those of quantum statistical mechanics \cite{nico2020}. 
Unlike other forms of ergodicity breaking, MBL is believed to be robust to Hamiltonian perturbations and therefore, constitute a genuine, stable non-equilibrium phase of matter. These systems generalize Anderson localization to interacting systems and are characterized by a number of interesting properties. 
They include the absence of 
particle transport~\cite{PhysRevB.90.174202}, 
a characteristic logarithmic growth of entanglement in 
time~\cite{PhysRevLett.109.017202,PhysRevB.77.064426}, 
highly excited states obeying 
an area law 
for entanglement entropies~\cite{AreaReview,Bauer,PhysRevB.91.081103,friesdorf_many-body_2015} and characteristic
multi-partite correlations
~\cite{PhysRevB.92.180202,PhysRevLett.118.016804},
 a memory of initial states
 in non-equilibrium dynamics~\cite{PhysRevB.90.174202,1408.5148,ngupta_Silva_Vengalattore_2011}, 
 and a number of other striking features. Because of these unique properties violating the 
 eigenstate thermalization hypothesis~\cite{MBLRMP}, 
 MBL as a phase of matter and the mechanisms leading to it are currently under study to stabilize and 
 explore other technological and physical applications such as quantum time crystals~\cite{WilczekTC,ElseTCPRL2016,YaoDTCPRL2017,ZhangTCNature2017,ChoiTCNature2017}.

Despite its importance and potential applications, the experimental evidence of MBL has so far basically exclusively 
been found in synthetic quantum systems including cold atomic setups~\cite{SchreiberMBLScience} and in 
superconducting circuits~\cite{RoushanMBLScience}. 
Both of these setups require precise control and the manipulation of individual atoms or 
qubits. At the same time, these approaches are limited to a relatively small fixed number of particles. 
This raises the question of whether properties of MBL systems can be experimentally probed outside such set-ups in real \emph{quantum materials}. Unlike the cold atomic set-ups, these systems are actual quantum many body systems in the  thermodynamic limit.
While possible signatures of MBL have been found in some real materials, the effect of electron-phonon coupling remains unclear in these materials~\cite{OvadiaSciRe2015}. 
Similar indirect evidence has been found in driven nuclear spin-chains~\cite{WeiPRL2018}, disordered super lattices~\cite{nguyen_signature_2020} and disordered magnets~\cite{SilevitchNatComm2019},  while conclusive results are lacking. 

\begin{figure}
    \centering
    \includegraphics[width=.95\linewidth]{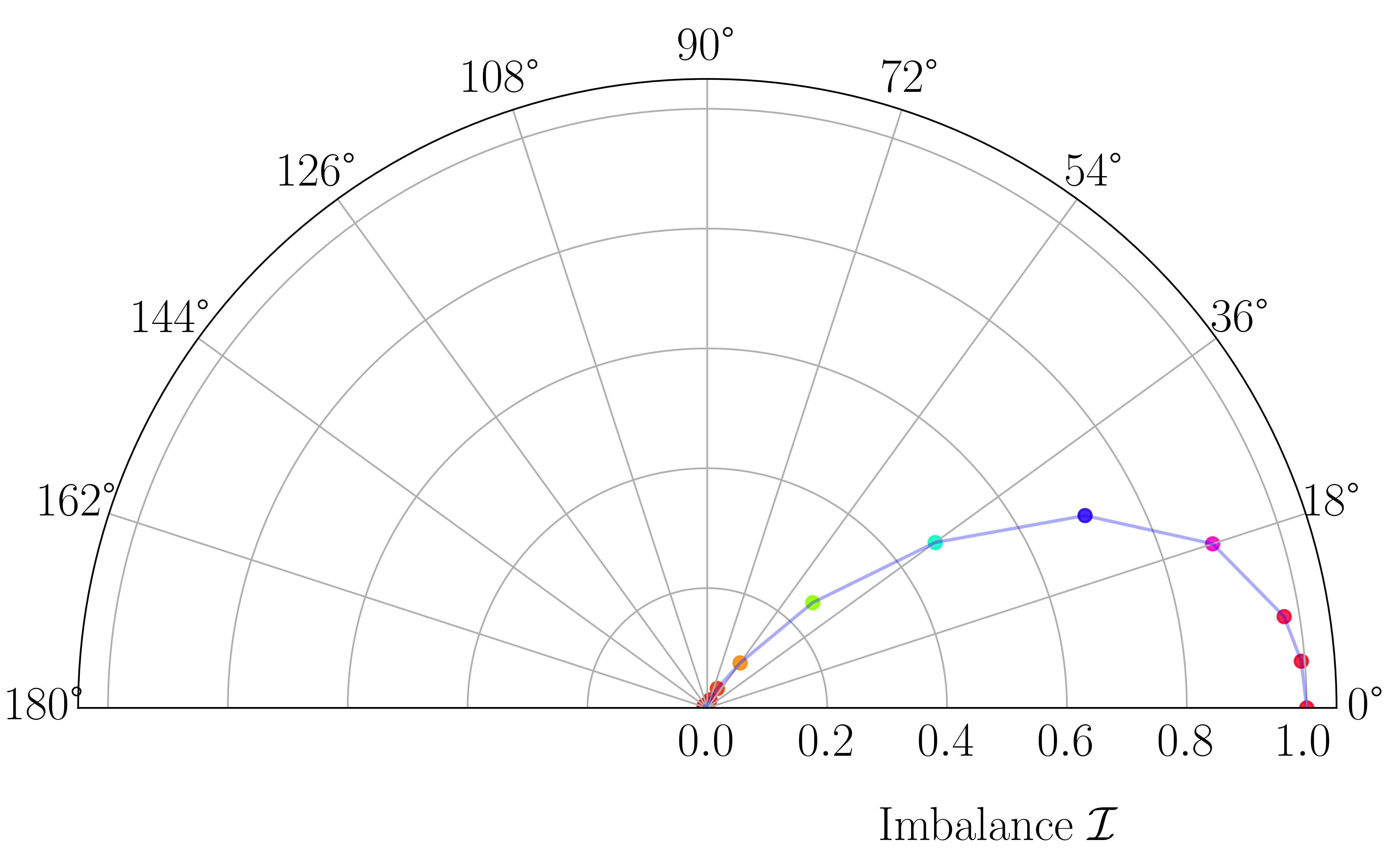}
    \caption{Particle imbalance on the radial axis as a measure of localization for the binary random bond XXZ model with doping strength $\delta=0.3$. $\theta$ interpolates between the random bond Ising model at $\theta=0^\circ$, the random bond Heisenberg model at $\theta=90^\circ$ and the random bond hopping model at $\theta=180^\circ$ (cf.~the last paragraph of 
    Section \ref{sec:analysis}). The localization of the random Ising model extends to non-zero values of $\theta$ and transits into the ergodic phase around $\theta\approx54^\circ$. 
    }
    \label{fig:radial_imbalance}
\end{figure}

In this work, 
we propose a route towards 
synthesizing actual materials exhibiting properties of 
many-body localization. We suggest that this can be 
done by combining two or more related, iso-structural compounds displaying spin chain magnetism which can be experimentally synthesized in the laboratory. We investigate in detail how different ratios of the compounds in the mixing lead to different localization behaviour, discuss the effects of the choice of the compounds and the impact of the coupling to phonons and other features and provide a detailed numerical analysis of the  different scenarios. Hence, our work is meant to act as a \emph{blueprint} of the experimental synthesis of real materials featuring properties of MBL. We complement this guide by an intuitive argument and numerical study concerning the intricate interplay between the electronic structure and the lattice phonons. 
In this way,
we suggest how signatures of MBL may be robust albeit the presence of phonons. What is more, this argument 
may lend a novel perspective to an open question about
the effective decoupling of the spin degrees of freedom from the phonon background \cite{MBLRMP, altshuler_jumps_2009, OvadiaSciRe2015}. 

\section{Theoretical set-up}

\begin{figure*}
    \centering
    \includegraphics[width=.9\linewidth]{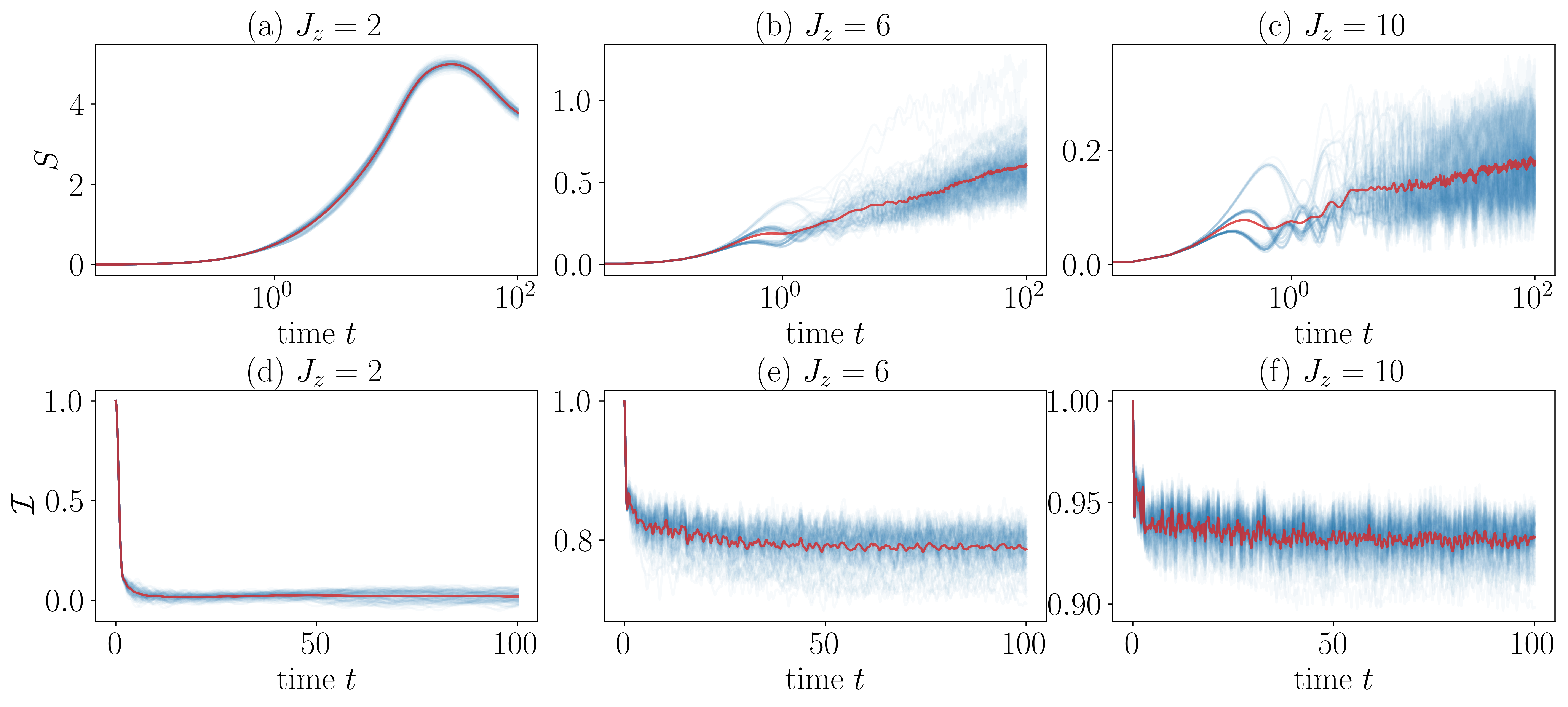}
    \caption{The top row shows the entanglement entropy $S$  as a function of time and the bottom row shows the particle imbalance $\imbalance(t)$ at doping strength $\delta=0.3$ for various disorder strengths $J_z=J_z^{(0)}$. Sub figures (a) and (d) correspond to disorder strength $J_z=2$, (b), (e) to disorder strength $J_z=6$ and (c), (f) to disorder strength $J_z=10$. Blue shaded regions correspond to the 100  disorder configurations, the disorder averages are plotted in red. 
    }
    \label{fig:doping0.3_various_eta_avg}
\end{figure*}

\subsection{Review and methods}
Many body localization naturally extends Anderson's idea of order from disorder from single particle to many body physics: Coupling the local degrees of freedom to a source of randomness in an adequate way, the ergodic behaviour of the time evolution of a quantum system is broken. The most studied model admitting many body localization is given by what we here call the \emph{standard MBL} 
(sMBL) Hamiltonian 
\begin{align}
    \HsMBL = &\sum_i \frac{J_\perp}{2}\left(s_i^+s_{i+1}^- + s_i^-s_{i+1}^+\right)\nonumber \\
    + &\sum_i\left(J_zs_i^zs_{i+1}^z + W h_i \sigma_i^z\right)\,,\label{eq:HsMBL}
\end{align}
where the real $W$ is the \emph{disorder strength} and $h_i\in [-1, 1]$ is a uniformly and identically distributed random 
variable. $s_i^x, s_i^y, s_i^z$ 
denote Pauli operators at 
site $i$,
while
$s_i^\pm := s_i^x \pm i s_i^y$.
$J_\perp$ is the hopping strength and $J_z$, the interaction strength of the magnetic coupling between neighbouring spins, on a spin chain. At $J_z\rightarrow0$ the model becomes Anderson localized for any finite value of $W$. While the localizing effects are destroyed by introducing a finite interaction $J_z\neq0$ at weak disorder strengths $W$, the system remains localized for sufficiently large values of $W$. Moreover, it is easy to see that the model becomes diagonal in the $z$ basis and that the quasi-local constraints of motion become the local $s_i^z$ operators in the limit $W\rightarrow\infty$. While $\HsMBL$ is defined with respect to a continuous disorder field, discrete disorder fields with only finite levels have also been studied~\cite{andraschko_purification_2014,Kshetrimayum2DMBLPRB2020,Kshetrimayum2DTCPRB2021}.

To proceed, in this work, we discuss effectively one-dimensional models. These
are captured by  \emph{matrix product states} 
\cite{DMRGWhite92,quant-ph/0608197} along with the \emph{time evolving block decimation} method for our simulations \cite{PhysRevLett.93.040502,Daley}, as implemented by the TeNPy library~\cite{tenpy}, that allow to keep track of non-equilibrium quantum systems for not too long times. 
The use of such
\emph{tensor network methods}
\cite{Orus-AnnPhys-2014,VerstraeteBig}
comes along with a number of advantages. They are extremely accurate in their realm of applicability.
What is more,
entanglement entropies
\cite{AreaReview} can be computed with little 
effort. 
This feature will be useful in determining whether a system exhibits 
properties of MBL or not as the entanglement entropy in many body localized systems displays 
an unbounded logarithmic growth in time \cite{PhysRevLett.109.017202,PhysRevB.77.064426},
starting with a product state that has no entanglement.
For a state vector $\ket{\psi}$, the 
\emph{entanglement entropy} 
with respect to a subsystem $A$ is given by the von-Neumann entropy $S(\rho) := -\Tr\rho\ln\rho$ of the reduced quantum state on $A$ as
\begin{align}
    S_A := S(\rho_A) 
    \intertext{with the reduced density matrix}
    \rho_A:=\Tr_{\backslash A}\ketbra{\psi}\,
\end{align}
where $\Tr_{\backslash A}$ is the partial trace on the complement of $A$.
Another key property of many body localized systems is the conservation of classical information that can be quantified by 
means of the classical information capacity
\cite{nico2020}
that in turn can be phrased in terms of a Holevo
quantity. In the case of $\HsMBL$ at strong disorder 
strength $W$, this coincides with the information in the computational basis which is captured by the 
particle imbalance. The \emph{particle imbalance} $\imbalance$ measures the amount of the residual,
conserved information in the computational basis and is defined as 
\begin{align}
    \imbalance(t) := -\frac{1}{n}\sum_{i=1}^n (-1)^i\expval{ s_i^z}{\psi(t)}
\end{align}
where $\ket{\psi(t)}=\exp(-itH)\ket{\psi_0}$ is the time evolved state vector, $\ket{\psi_0}=\ket{\uparrow,\downarrow,\uparrow, \dots, \uparrow,\downarrow}$ is the Néel state vector 
and $n$ is the size of the chain. 
The \emph{quasi-local constraints of motion} (qLCOM) 
 \cite{PhysRevB.90.174202,PhysRevLett.111.127201}
of the many body localized fixed point $\HsMBL(W\rightarrow\infty)$ are the local $s^z$ operators. Therefore, the particle imbalance $\imbalance(t)$ is a constant of motion for $W\rightarrow\infty$. For finite $W$, it is a measure of the extent to which local information is preserved. Hence, as the imbalance relaxes to zero for ergodic systems, it is an indicator for how far from this fixed point the system is. The combination of the preserved particle imbalance and logarithmic growth of entanglement highlights core features of MBL as opposed to Anderson localization: While the classical information which is in our case measured with respect to the computational basis and is indicated by the particle imbalance is preserved in many-body localized systems as well as in Anderson localized systems in time, the quantum correlations as indicated by the entanglement entropy may still propagate slowly through the system, while they are preserved in Anderson localized systems. Indeed, while MBL inhibits particle transport, slow quantum information propagation
still takes place \cite{1412.5605,PhysRevB.96.174201}.

\subsection{Doping induced interaction disorder}
We now move to discussing the core topic of this work, which concerns ways of
synthesizing and investigating real materials featuring properties of MBL. Our premises 
are the following:
Firstly, we will consider 
\emph{discrete disorder} featuring a finite number of levels only rather than the continuous disorder strength, $h_i$ of $H_{sMBL}$ as
specified in Eq.~(\ref{eq:HsMBL}). Secondly, instead of investigating on site disorder, we will consider disorder directly on the interactions which can be achieved by doping on non-magnetic sites of magnetic spin chain compounds. While we believe that this model is closer to experimental realizations, the analysis does carry over qualitatively from on site disorder since both models have the same MBL fixed point. And thirdly, we move away from uniformly independently distributed disorder and investigate disorder distributions that are mimicking the doping process of the real materials more closely and are hence correlated. 

We aim to exploit the inherent randomness in the \emph{doping process} of two or more iso-structural spin chain compounds which have related but different magnetic Hamiltonians as a source of randomness in the system. 
We consider two compounds labelled by $0$ and $1$ both described by an effective XXZ interaction
\begin{align}
    \HXXZ^{(j)} = &\sum_i \frac{J_\perp^{(j)}}{2} (s_i^+s_{i+1}^- + s_i^-s_{i+1}^+) 
    + \sum_i J_z^{(j)} s_i^zs_{i+1}^z
\end{align}
where $j=0, 1$ denotes the respective compound whose hopping strength is $J_\perp^{(j)}$ and interaction strength is $J_z^{(j)}$. 
The real values of $J_\perp^{(j)}$ and $J_z^{(j)}$ are fixed by the choice of the compounds, they cannot be arbitrarily tuned since in reality only a few compounds will form in the required structure and can be doped into each other.
If we dope these two materials, the resulting Hamiltonian will consist of strings of couplings of random length stemming from material $0$ alternating with strings of couplings stemming from material $1$. Hence, to a reasonably good approximation, 
the resulting Hamiltonian, $\Hdoping$, will be of the form 
\begin{align}\label{h_doping}
    \Hdoping = &\sum_i \frac{J_\perp^{c(i)}}{2} (s_i^+s_{i+1}^- + s_i^-s_{i+1}^+) 
    + \sum_i J_z^{c(i)} s_i^zs_{i+1}^z
\end{align}
where $J_\perp^{c(i)}$ and $J_z^{c(i)}$ are distributed randomly according to the doping strength $\delta$, which gives the ratio of the two materials. This is reflected in the configuration $c$, where $c\in\{0,1\}^{\times (n-1)}$ is a length $n-1$ bit string with one bit $c(i)$ for each bond $i$
denoting the bond values due to the doping process. A fraction $\delta$ of these $n-1$ bonds have coupling values $J_\perp^{(1)}$ and $J_z^{(1)}$ while the remaining fraction
$1-\delta$ of $n-1$ 
bonds have $J_\perp^{(0)}$ and $J_z^{(0)}$.

In the laboratory, one has a very precise control over the doping strength. In fact, this is the parameter with the highest experimental control. Therefore, it is crucial to take it into the analysis and keep it as a separate parameter that is able to influence the stability of the MBL state. 
Accordingly, one has to sacrifice the independence of the individual bonds in the sampling process. 
To see this consider the following situation: if the doping strength is set to $\delta=0.5$ and the first $n/2$ bits have been assigned the values $c_i=1$ for all $i=1$ through $n/2$ then the remaining bits must be assigned the value $0$. Hence, the individual bits are not distributed independently of each other. 
More generally, the disorder can be sampled according to the following algorithm.

\begin{algorithm}[H]
\caption{Sample doping configuration}
\begin{algorithmic}
\REQUIRE System size $n \geq 2$ and doping strength $ \delta \in[0,1]$
\STATE $c \leftarrow (0,\dots, 0)\in\{0,1\}^{n-1}$
\STATE Sample $N=\lceil \delta (n-1)\rceil$ many distinct positions $i\in[n-1]$ and denote the set by $c^{(1)}$
\FOR{$i\in c^{(1)}$} 
    \STATE $c_i\leftarrow1$
\ENDFOR
\RETURN $c$
\end{algorithmic}
\end{algorithm}

\begin{figure*}
    \centering
    \includegraphics[width=0.9\linewidth]{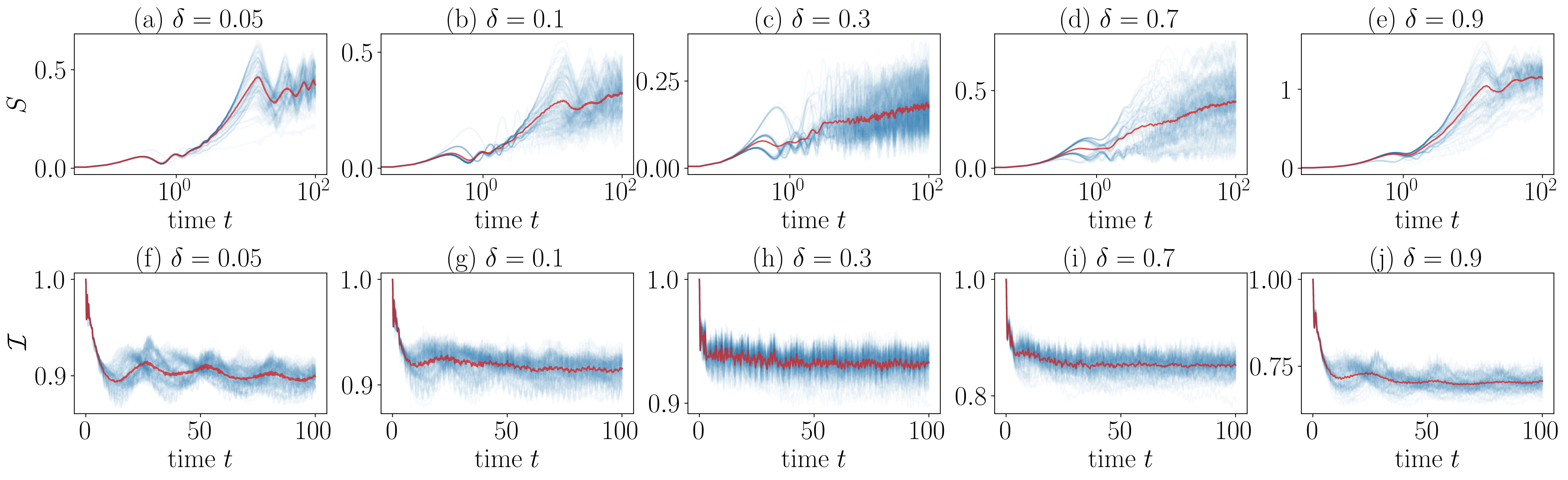}
    \caption{In all plots $J_z^{(0)}=10$ and $J_z^{(1)}=5$ with homogeneous $J_\perp=1$. 
    The top row shows the entanglement entropy $S$ as a function of time and the bottom row shows the particle imbalance $\mathcal{I}$ as a function of time for doping strengths $\delta=0.05,\dots,0.9$. 
    The blue shaded regions are the plots of the individual runs of the $100$ disorder configurations and the red plot is the ensemble average. 
    }
    \label{fig:eta10_various_delta_avg}
\end{figure*}

\section{Analysis}\label{sec:analysis}
Let us first consider a mean field approximation to establish that the disorder induced by doping indeed has a non-trivial effect. To this end, we observe that the Hamiltonian $\Hdoping$ in Eq.~(\ref{h_doping}) has a mean field approximation with respect to the initial computational basis state vector
$\ket{\psi_0}=\ket{j_0,\dots,j_n}$
with $j_i\in\{\uparrow, \downarrow\}$
\begin{align}
H_{\text{doping}}^{\text{MF}} = \sum_i& \frac{J_\perp^{c(i)}}{2}(s_i^+s_{i+1}^-+s_i^-s_{i+1}^+) 
+\sum_i \sigma(j_{i+1}) J_z^{c(i)} s_i^z\nonumber
\end{align}
where $\sigma(\uparrow)=1$ and $\sigma(\downarrow)=-1$ quantifies the local magnetization. In particular, for homogeneous $J_\perp=J^{(0)}_\perp=J^{(1)}_\perp$ and disordered $J_z$($J^{(0)}_z \neq J^{(1)}_z$), 
we observe that the mean field Hamiltonian resembles the non-interacting Anderson model for localization lending a first evidence for the non-trivial effect of interaction disorder. 

To obtain evidence for the existence of a many body localized phase for the Hamiltonian $\Hdoping$ mimicking doped materials, 
we simulate the time evolution of the Hamiltonian $\Hdoping$ in Eq.~(\ref{h_doping}) 
using matrix product states and the TEBD algorithm from TeNPy~\cite{tenpy} for different system sizes up to $L=64$ and bond dimensions up to $\chi=1024$ for different parameters which we discuss below. We compute the particle imbalance and the entanglement entropy for the setting where we fix the hopping strength $J_\perp^{(i)}=1$ for both $i=0,1$ 
while the ratio of the interaction strength was fixed at $J_z^{(0)}/J_z^{(1)}=2$. The value of the $J_z^{(0)}$ can then thought of as the disorder strength and can be varied along with the doping strength $\delta$ which parameterizes the fraction of material labelled $1$ in the compound. While the choice of these parameters may appear arbitrary, we would like to highlight that the purpose of our work is to show that the disorder induced in the interaction (bond) of the Hamiltonian can lead to signatures of MBL. We have also done simulations for the disorder induced in the hopping strengths of the Hamiltonian but we have found that this does not play a significant role in the localization 
(c.f.~Fig.~\ref{fig:radial_imbalance}). We have also assumed that the interaction strength of one of the compounds is twice that of the other. 
This is motivated by the observations that (1) both interaction strengths should depend on the disorder strength and (2) both interaction strengths must be sufficiently different to distinguish from a homogeneous strong Ising anisotropic material. Moreover, as the microscopic structure of the materials should be similar for the model of the doping to be justified, we chose a factor of two such that the discrepancy is not too large. 
We have found that, for fixed doping strength, the bigger the difference in the interaction strength of the two materials, the stronger the localization (data shown in Appendix \ref{app:interaction-gap}).

After fixing the doping strength to $\delta=0.3$ and increasing the disorder strength $J_z^{(0)}$, we find a transition from an ergodic phase for $J_z^{(0)}\lesssim4$ into a localized phase for $J_z^{(0)}\gtrsim4$. As shown in Fig.~\ref{fig:doping0.3_various_eta_avg}, for weak disorder strength $J_z^{(0)}=2$, the entanglement entropy grows approximately linearly until it starts oscillating around its saturation value, while the imbalance immediately drops close to its equilibrium value $\imbalance_\infty\approx0$. Increasing the disorder strength $J_z^{(0)}=6$ above the transition point we find logarithmic growth of entanglement typical of MBL while the imbalance saturates around a finite value $\imbalance_\infty\approx0.8$. Increasing the disorder strength further $J_z^{(0)}=10$ only strengthens the MBL phenomenology: The entropy grows slower while still asymptotically logarithmic and the saturation value for the imbalance $\imbalance_\infty\approx0.93$ moves closer to one such that we conclude to find evidence for MBL. 

The analysis of the effect of the doping strength $\delta$ turns out to be more subtle. 
In particular, we could not detect a definite doping strength for which the MBL phenomenology breaks down.
However, due to the sparse occurrence of doped sites in the weak doping regime $\delta<0.1$ and $\delta>0.9$ large system sizes and a large bond dimension are required to obtain meaningful results and to avoid finite size or finite entanglement effects as shown in Fig.~\ref{fig:eta10_various_delta_avg}, we have investigated the dynamical behaviour of the system for $\delta=0.05,\dots,0.9$ in terms of the entanglement entropy and the particle imbalance. Starting from $\delta=0.05$ we find that the localizing behaviour witnessed by the entanglement entropy and the particle imbalance increases upon increasing $\delta$. However, a peak is reached around $\delta=0.3$ and the behaviour becomes less pronounced upon increasing $\delta$ beyond $\delta=0.5$ (data not shown).
This can be explained by the dominance of $J_z^{(0)}$: While larger doping strengths $0.6>\delta>0.3$ may induce a more complex pattern into the doping configuration $c$ leading to stronger interference effects, the norm of the unperturbed Hamiltonian corresponding to the terms diagonal in the $z$ basis only decreases as more terms stemming from material 1 which carry a weight $J_z^{(1)}=J_z^{0}/2$ are present. This explains why the localizing effect is far from being symmetric around the doping strength centre  $\delta=0.5$. Because the dominant fraction of terms carries a weight $J_z^{(1)}$ if $\delta>0.5$, the full phenomenology is comparable to some $\delta<0.5$ at a decreased disorder strength $\tilde J_z^{(0)}$.  
Fig.~\ref{fig:eta10_various_delta_avg} (a), (b), (f) and (g) (respectively (e) and (j)) display the effects of the entanglement entropy and particle imbalance in the weak (respectively strong) doping regime. 

Interestingly, the particle imbalance does seem to saturate to a finite value despite the very weak doping strengths $\delta=0.05$ and $\delta=0.1$ (respectively $\delta=0.9$). Similarly, the entanglement entropy does seem to grow at most logarithmically. 
However, we note that the particle imbalance has a smoother transition into the saturated regime compared to the intermediate doping strengths. Similarly the entanglement entropy seems to be saturated at a finite value for the weak doping strength $\delta=0.05$ and has a smaller variance compared to the intermediate doping strengths.
This behaviour is very reminiscent of the undoped XXZ model with a strong Ising anisotropy for finite system sizes. Within our numerical setting we, therefore, find evidence for MBL only for doping strengths $0.9\gtrsim\delta\gtrsim.1$. We note that, while we do not have any evidence for that, the localizing phenomenology at extremal doping strengths may still reestablish at sufficiently large system sizes.

While doped materials will have different interaction strengths, they will realistically also have varying hopping strength. Therefore, we have analyzed the effect of material dependent hopping terms
$J_\perp^{(i)}=J_\perp\pm\epsilon$ onto the MBL phenomenology. 
We have found that the imbalance and the entanglement entropy behave in exactly the same manner as in the case of homogeneous hopping terms (data not shown). Thus we conclude that disorder in the hopping has no observable effect on the MBL phenomenology.

In order to emphasize the importance of a strong anisotropy leaning towards a very Ising like material, also dealing as a intuition for the material selection for experimentalists, we have analyzed the Hamiltonian $\Hdoping$ (\ref{h_doping}) with the $J_z^{(i)}$'s and $J_\perp^{(i)}$'s varying between random interaction ($J_\perp^{(i)}=0$, $J_z^{(i)}\neq0$) through random Heisenberg ($J_\perp^{(i)}=J_z^{(i)}$) to random hopping ($J_\perp^{(i)}\neq0$, $J_z^{(i)}=0$) Hamiltonians. We set $J_z^{(0)}=2J_z^{(1)}=\cos(\theta)$ and $J_\perp^{(0)}=2J_\perp^{(1)}=\sin(\theta)$ and fixed $\delta=0.3$. 
Thus, $\theta=0^\circ$ corresponds to the random Ising point, $\theta=45^\circ$ corresponds to the random Heisenberg point and $\theta=90^\circ$ corresponds to the random hopping point.
As shown in Fig.~\ref{fig:radial_imbalance}, we find that strong disorder alone is not sufficient to lead to localization but that the disorder must be strong in a local basis such as the computational basis. This is in good agreement with the theoretical understanding of MBL \cite{Imbrie2016}.

\section{Experimental set-up}

\begin{figure}
\centering
\includegraphics[width=.9\linewidth]{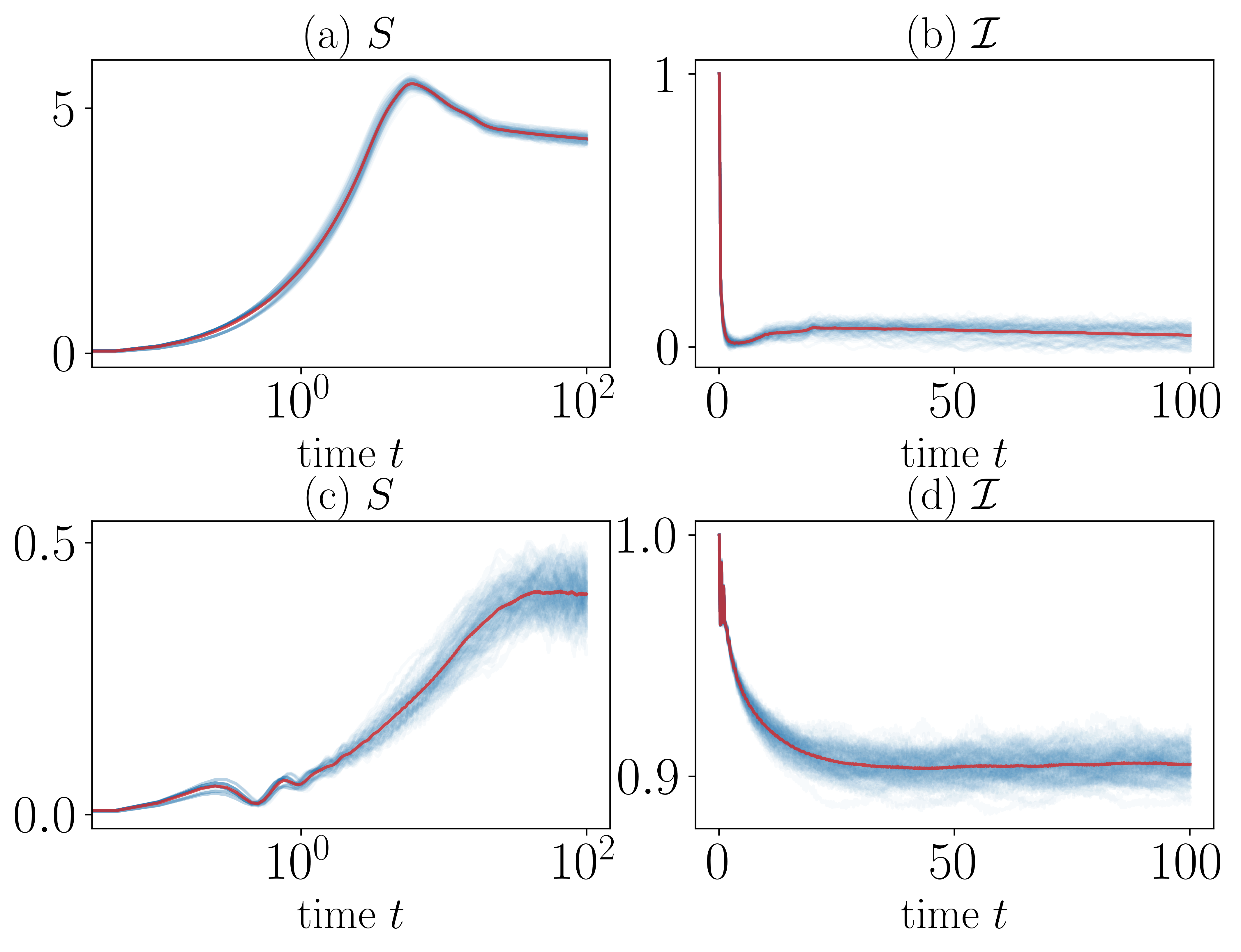}
\caption{Numerical results for the compounds of real materials with doping strength $\delta=0.3$. (a) and (b) shows the results for the compound made of SrCo$_2$V$_2$O$_8$ and BaCo$_2$V$_2$O$_8$ while (c) and (d) shows the results for the compound made of CsCoCl$_3$ and CsCoBr$_3$. (a) and (c) show the entanglement entropy. (b) and (d) show the particle imbalance. Blue shaded regions correspond to the 100 disorder configurations and the red line the corresponding average.
}
\label{fig:CoVO}
\end{figure}

As we have seen in the previous section, in order to obtain disorder induced MBL it is crucial to combine spin chain compounds that both come with a strong Ising-like anisotropy while at the same time being sufficiently different in terms of their $J_z$ coupling strength. From an experimental perspective a further constraint is that 
the two materials in question must be sufficiently similar in terms of their crystal structure and space group so that the structural properties such as lattice parameter and atomic positions vary smoothly with doping, preserving the spin chain characteristics
so that the resulting material is indeed composed of alternating strings of compound 0 and 1. This already eradicates a number of promising candidate pairs which, 
while well matched in terms of their magnetic Hamiltonians, fail to be sufficiently similar on a microscopic level.

In this section, we are going to compute the same quantities of entanglement entropy and particle imbalance as in the previous section for two candidate pairs of compounds. 
In both cases the two constituent compounds are chosen to have very similar or identical space groups and their chemical formulas differ only by one ion. This ion is not the magnetic ion that forms the spin chain but rather a non-magnetic ion that influences the intra-chain exchange interactions by slight alterations of the magnetic exchange pathway and the spacing of the magnetic ions. While both pairs will fail to admit a clear evidence of MBL we include this analysis as it provides an intuition for the material selection and the underlying phenomenology. 

The first quantum material we investigate is the compound Sr$_{1-\delta}$Ba$_\delta$Co$_2$V$_2$O$_8$ made from the parent compounds SrCo$_2$V$_2$O$_8$ and BaCo$_2$V$_2$O$_8$ which contain screw-chains of magnetic Co$^{2+}$ ions that have effective spin-1/2. As shown in Refs.~\cite{bera_spinon_2017,PhysRevLett.114.017201,PhysRevLett.115.119902}, both materials have very similar microscopic structures and correspond to anisotropic Heisenberg spin-chain anti-ferromagnets. Here, SrCo$_2$V$_2$O$_8$ will constitute material 0 and can be described by the coupling strengths $J_z^{(0)}=7\text{meV}$ and $J_\perp^{(0)}=3.92\text{meV}$ \cite{bera_spinon_2017}. Likewise BaCo$_2$V$_2$O$_8$ constitutes material 1 with coupling strengths $J_z^{(1)}=5.8\text{meV}$ and $J_\perp^{(1)}=3.074\text{meV}$ \cite{PhysRevLett.114.017201,PhysRevLett.115.119902}.
As shown in Fig.~\ref{fig:CoVO} (a) and (b), this combination quickly thermalizes. In particular, the particle imbalance immediately drops to zero while the entanglement entropy grows linearly in time. Both clear indications for the absence of ergodicity breaking. As the ratio between the interaction and the hopping terms is only of the order $J_z^{(0)}/J_\perp^{(0)}\approx1.8$, this comes at no surprise as the Ising anisotropy is not yet sufficiently pronounced 
(c.f.~Fig.~\ref{fig:doping0.3_various_eta_avg}).

We observe the other extreme of non-localized behaviour for the compound CsCoCl$_{3(1-\delta)}$Br$_{3\delta}$, made of the Ising-Heisenberg anti-ferromagnets CsCoCl$_3$ and CsCoBr$_3$ \cite{nagler_ising-like_1983} where the anisotropic spin chain magnetism also arises from Co$^{2+}$ ions. The couplings of the effective Hamiltonian are $J_z^{(0)}=13.41\text{meV}$ and $J_\perp^{(0)}=1.314\text{meV}$ corresponding to CsCoBr$_3$ and $J_z^{(1)}=11.78\text{meV}$ and $J_\perp^{(1)}=1.008\text{meV}$ for CsCoCl$_3$ respectively. Fig.~\ref{fig:CoVO} (c) and (d) show the entanglement entropy and particle imbalance, respectively, of this material. In contrast to the previous material, we observe a saturation of the entanglement entropy at a small finite value as well as a saturation of the particle imbalance. While the saturation of the imbalance seems to indicate localizing behaviour the saturation of the entanglement entropy rather hints at a finite size effect (noting that the results are already converged in the bond dimension). While uncommon for many body localized systems, this finite size effect is generic for clean systems admitting a strong Ising anisotropy at finite system sizes. Moreover, in contrast to the results shown in Fig.~\ref{fig:doping0.3_various_eta_avg}, we observe only a small variance of the individual configurations around the their mean in 
Fig.~\ref{fig:CoVO} and the transition into the saturated regime seems less sharp. 
We, therefore, suspect the observed saturations of the entanglement entropy and the particle imbalance shown in the plots do not witness MBL but resemble finite size effects induced by the strongly pronounced Ising anisotropy of each of the materials. Moreover, we suspect that the small variance is due to the small difference of the respective coupling terms $J_z^{(0)}-J_z^{(1)}\approx1.35\text{meV}$ such that the individual compounds look almost homogeneous and only admit minor differences in their evolution 
(c.f.~Appendices~\ref{app:interaction-gap} and \ref{app:finite-size-scaling}). 

We conclude this section with the following observations already elaborated on in Section~\ref{sec:analysis}. In order to achieve many body localization in a compound of two materials it is crucial that both materials are strongly leaning towards an Ising anisotropy. At the same time, however, both materials need to be significantly different in the strength of their Ising coupling. We presented two possible combinations of different materials, each of which only fulfills one of the two conditions, and observed that neither of them does give rise to many body localization. Therefore, we believe that the major challenge in the search for a compound that admits MBL lies in the trade-off between the similarity of their microscopic structure versus their pronounced difference in terms of the strength of the (dominating) Ising coupling. While the first is necessary to ensure that our doping-toy model is a good effective description, the latter is, as shown above, crucial for the actual MBL phenomenology.

\section{Protecting MBL from phonons}
It is generally believed that the phonons in a real material constitute a thermal bath and that the 
non-zero electron-phonon coupling will lead to thermalization of the system \cite{MBLRMP,Agarwal,Eckardt}. For this reason, this 
aspect requires particular attention when devising a blueprint for realizing MBL in real materials.
In this section, we will reinvestigate the microscopic description of the electron-phonon system in order to derive a fresh perspective on the robustness of the localizing phenomenology. In particular this analysis lends guidance towards  engineering materials in which the underlying disorder can be used to control and identify the artefacts of the induced localization. 

To begin with let us recap the microscopic model, where without loss of generality we directly work in the spin rather than in the electron picture. The Hamiltonian can be recast into the form
\begin{align}
    H = H_{\text{spin}} + H_{\text{phonon}} + H_{\text{spin-phonon}}.
\end{align}
Here $H_{\text{spin}}$ is the Hamiltonian describing the microscopic spin degrees of freedom. The microscopic phonon Hamiltonian is described by a system of harmonic oscillators representing the free bosons
\begin{align}
    H_{\text{phonon}} = \sum_{\langle i,j\rangle} \omega_{i,j} a_{i,j}^\dagger a_{i,j} + \sum_{\substack{\langle \langle i,j\rangle,\langle k,l\rangle\rangle}} \kappa_{i,j,k,l}a_{i,j}^\dagger a_{k,l},
\end{align}
where every bond of the lattice is identified with an oscillator with real energy $\omega_{i,j}$ and neighbouring oscillators can exchange excitations via a hopping term $\kappa_{i,j,k,l}$. Note that we use double indices referring to the bonds in order to align with the notation of the spin Hamiltonian. We model the interaction between spin and phonon degrees of freedom as
\begin{align}\label{def:spin-phonon}
    H_{\text{spin-phonon}} = \sum_{\langle i,j\rangle} g_{i,j} \left(a_{i,j}^\dagger + a_{i,j}\right) \left(s_i^+s_j^- + s_i^-s_j^+\right).
\end{align}
This is, a hopping of a spin excitation over the bond $\langle i,j\rangle$ couples to the position operator $X_{i,j}\propto (a_{i,j} + a_{i,j}^\dagger)$ corresponding to the respective oscillator such that the lattice vibrations modulate the spin transport according to the coupling $g_{i,j}$. For a homogeneous material $\omega_{i,j}=\omega$ and $\kappa_{i,j,k,l}=\kappa$ such that $H_{\text{phonon}}$ is diagonal in the momentum basis and the phononic degrees of freedom can be expected to form a thermal bath. Because the phonon dynamics is in general faster than the electronic dynamics this gives rise to an effective heat bath to which the spin or electric degrees of freedom couple through the interaction term. The corresponding coupling term $g$ has a finite material dependent value such that the spin degrees of freedom  are believed to thermalize in the long run. 

While this analysis gives good results for homogeneous lattices we argue that it must break down for disordered systems and at sufficiently low temperatures and sufficiently strong disorder. In addition, the following argument may give a microscopic explanation to the effective decoupling between electronic and phononic degrees of freedom as laid out in, e.g., 
Ref.~\cite{OvadiaSciRe2015}.

\subsection{Isolated phonons}
We start by observing the effect of the disorder on the microscopic phonon Hamiltonian. First we note that for a disordered material 
the oscillator energies $\omega_{i,j}$ will be not homogeneous, but distributed according to the bond-disorder. Hence, the isolated phonon Hamiltonian  $H_\text{phonon}$ of one dimensional systems will be Anderson localized where the localization length decreases with the underlying disorder strength \cite{anderson_absence_1958, abrahams_50_2010}. 
The disorder in the phonon Hamiltonian is also
responsible for the inaptness of the argument made in Ref.~\cite{PhysRevB.68.134207}, which
 suggest a divergence of the localization length at low
frequencies.

This case is particularly interesting in three (and higher) dimensions. While there is no generic Andersson localization in three dimensions, there is a disorder dependent mobility edge. States below some critical energy $E_c(\Delta)$  will be localized while states above this threshold behave in an ergodic fashion. 
Moreover, the mobility edge grows with the disorder strength $\Delta$ where the exact dependence depends on the specific system \cite{abrahams_scaling_1979, kondov_three-dimensional_2011, licciardello_conductivity_1978}. 

In contrast, two dimensional systems represent a marginal dimension between the one and three dimensional phenomenology. While all states throughout the energy spectrum are localized, the localization length of the phonons depends on their energy in  the following way. Below some critical energy scale $E(\Delta)$ which depends on the disorder strength $\Delta$ the system is localized exponentially well. Above this critical value, the localization length depends logarithmically on the energy. Hence, for most experimental set-ups this will resemble a mobility edge similar to the three dimensional case fairly well \cite{abrahams_scaling_1979, licciardello_conductivity_1978, white_observation_2020}.

\subsection{One dimensional systems}\label{sec:one-d}

The Anderson localization of the phononic degrees of freedom comes in nicely to the study of MBL with phonons. In truly one dimensional systems the electronic as well as the phononic degrees of freedom are localized. Hence, assuming a sufficiently short localization length in both systems, translating to a sufficiently large disorder strength, the interaction between the two systems is unlikely to induce any transport and we expect the electronic properties of the full system to be in qualitative agreement with the properties of the pure electronic system. In particular we believe that the compound system will remain many body localized.

A rigorous proof of this claim is beyond the scope of this work, but we would like to point out that the intuitive counter-argument of the phonons forming a heat bath is no longer valid in the present set-up. Moreover, as both, the spin and the phonon system are localized by the same disorder, we believe that the compound system is described by an MBL-like Hamiltonian in the following way.

The localization of electrons by MBL is believed to be robust to weak local Hamiltonian perturbations. The Anderson localized phase of the phonons is thought to become many body localized in the presence of couplings sufficiently weak compared to the disorder strength and is hence also believed to be robust. This can be seen
as evidence that this stability can extend to finite coupling strengths that  describe the electron-phonon interaction in one dimensional systems with phonons for  disorder strengths sufficiently strong compared to the electron-phonon coupling.
Finally, we would like to point out that since the phononic degrees of freedom alone do not thermalize, falsifying our claim would also give new insights into the mechanisms for reversing the ergodicity breaking introduced by the disorder.

\begin{figure*}
    \centering
    \includegraphics[width=.9\linewidth]{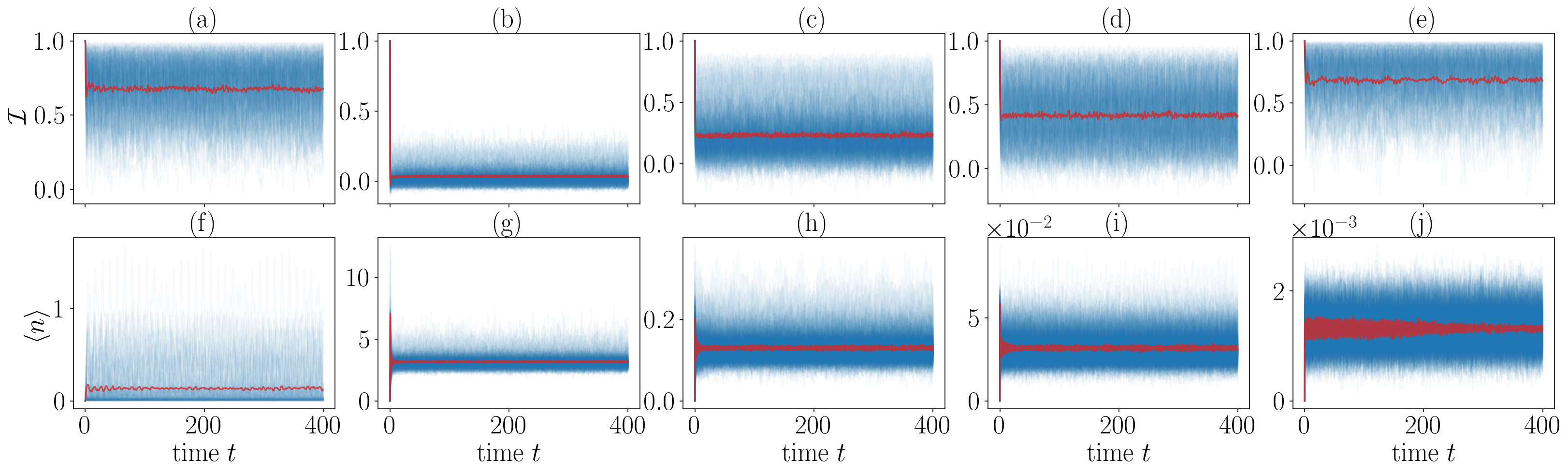}
    \caption{Imbalance $\mathcal{I}$ (a) - (e) and phonon occupation number $\expval{n}$ (f) - (j) time series simulated for $8$ spins and $p=1$ phonon mode with cut-of at $k=2^5=32$ levels. The spin Hamiltonian $H_\text{spin}$ is the Heisenberg model with random onsite potential with disorder strength $W=10$. The plots are for 100 disorder realizations, red curves are the corresponding averages. The phonon energy $\widetilde{\omega}$ and the spin-phonon coupling $g$ are given by $\widetilde{\omega}=10$ and $g=0.3$ for (a) and (f); $\widetilde{\omega}=10$ and $g=5$ for (b) and (g); $\widetilde{\omega}=50$ and $g=5$ for (c) and (h); $\widetilde{\omega}=100$ and $g=5$ for (d) and (i) and $\widetilde{\omega}=500$ and $g=5$ for (e) and (j). While the the system in the weak coupling limit (a) and (f) remains localized it becomes smeared out as $g$ is increased in (b) and (g). The system is brought back to a localizing dynamics as the phonon energy $\widetilde{\omega}$ is increased sufficiently as in (c) and (h) through (e) and (j). Note that the occupation number in (i) and (j) is re-scaled by $10^{-2}$ and $10^{-3}$, respectively.
    }
    \label{fig:phonon-coupling}
\end{figure*}

\subsection{Three dimensional systems}\label{sec:three-d}

The analogous chain of thought is less straight-forward for one dimensional spin systems embedded in a three dimensional material. 
The localization length of the phononic degrees of freedom with energies above the mobility edge $E_c(\Delta)$ will diverge.
The high energy sector of phonons therefore can be qualitatively treated similar to a homogeneous lattice, borrowing the equilibrating behavior. 
At low energies, sufficiently smaller than $E_c(\Delta)$, however, most phonons are well localized with short localization length. 
Hence, with the same argument as in Section~\ref{sec:one-d} those low energy phonons will not induce any transport.
Only the high energy phonons bias the system towards equilibration. To derive a qualitative toy model, we neglect the low energy well localized phonons with energies $e<\widetilde{\omega}$ below a critical energy $\widetilde\omega$ sufficiently below the mobility edge. Note that the critical energy may not be defined by the mobility edge but by a qualitatively too large localization length when compared to the electronic length scales. 
Hence, we rewrite the phononic and the interaction Hamiltonians as
\begin{equation}\label{def:model-phonon}
    \widetilde{H}_{\text{phonon}} = \sum_{e >\widetilde{\omega}}\widetilde{\omega}_e a_e^\dagger a_e
\end{equation}
and 
\begin{equation}\label{def:model-spin-phonon}
    \widetilde{H}_{\text{spin-phonon}} = \sum_{e >\widetilde{\omega}}\sum_{\langle i,j\rangle} \widetilde{g}_{e} \left(a_{e}^\dagger + a_{e}\right) \left(s_i^+s_j^- + s_i^-s_j^+\right).
\end{equation}
Where we also introduce a toy model for the spin-phonon coupling: Because the high energy part of the phonon Hamiltonian is approximately diagonal in the momentum basis a more rigorous treatment would first Fourier transform Eq.~\eqref{def:spin-phonon} and then truncate the low momentum modes. 
While this would lead to a more complicated form of the Hamiltonian if written in real space, the toy model as defined in Eq.~\eqref{def:model-spin-phonon} can be expected to have the same equilibrating potential. 
In fact, because the latter Hamiltonian couples all modes to all spins in a geometrically non-local fashion we believe that it even amplifies the potential equilibration.

Due to the temperature dependent energy distribution we can treat the corresponding spin-phonon Hamiltonian with an additional high energy cut of such that only the phonons from within a finite energy band contribute.
This interplay between the temperature dependent distribution of energy and energy dependent localization length leads to a (diverging) suppression of the phononic occupation number and hence an effective decoupling of the electronic from the phononic degrees of freedom at sufficiently low temperatures.
For very low temperatures and large disorder strength we can neglect the phononic coupling all along. 
Moreover, as our analysis in Section \ref{sec:numerical-phonon}  indicates, when $\widetilde{\omega}$ is sufficiently large compared to $\widetilde{g}$ the creation of phonons is strongly suppressed. Thus, in this regime 
the system will not contain enough high energy phonons to thermalize the spin degrees of freedom. 
We thus believe that traces of the MBL phenomenology will endure this regime.
This is phenomenologically in good agreement with experiments made in Ref.~\cite{nguyen_signature_2020}.

\subsection{Two dimensional systems}
Similar as for Anderson localization, the behavior of one dimensional spin systems embedded in two dimensional lattices is effectively the same as for those systems embedded in three dimensional lattices. 
The analysis is thus the same as in Section~\ref{sec:three-d}, only that the mobility edge $E_c(\Delta)$ is replaced by the critical energy at which the localization length transits from a constant to a logarithmic energy dependence.
However we do not expect any major differences in the phenomenology. 
This is in qualitative agreement with thin film experiments such as  Refs.~\cite{ovadia_electron-phonon_2009,OvadiaSciRe2015}.

\subsection{Numerical analysis}\label{sec:numerical-phonon}

The high energy sector of the three dimensional system is unlikely to be localized. According to our previous analysis, however, the full system may be non-ergodic due to the localization of the low energy sector. Hence, we expect insulating effects due to the localization to be visible in experiment in low to intermediate energy regions similar as in Ref.~\cite{nguyen_signature_2020}.
We conclude this section by a numerical small scale simulation via exact diagonalization to confirm this intuition. 
To this end we analyze Eq.~(\ref{toy-model-Hamiltonian}). 
As we find evidence for MBL in the doping Hamiltonian as defined in Eq.~(\ref{h_doping}) 
only for sufficiently large systems, we use the well established standard MBL Hamiltonian $H_{\text{spin}}=\HsMBL$ as in 
Eq.~(\ref{eq:HsMBL}).  
As such, we simulate the model Hamiltonian
\begin{equation}\label{toy-model-Hamiltonian}
    H_{\text{model}}^{p, k}(W,\widetilde{\omega}, \widetilde{g}) =  \HsMBL(W) + \widetilde{H}_{\text{phonon}}(\widetilde{\omega}) + \widetilde{H}_{\text{spin-phonon}}(\widetilde{g})
\end{equation}
where we have set the Heisenberg coupling parameters $J_z=J_\perp=1$.  $W$ denotes the disorder strength and the phononic degrees of freedom are parametrized by $q$ denoting the modes we include and $k$ the truncation parameter of each bosonic Hilbert space. 
Due to the numerical resources we have to trade off between $k$ and $q$. The delocalizing effects of the phonon bath should be already present in a bath consisting of a single phonon mode, i.e., $q=1$. Moreover, the dimensionality $k$ of the bath controls the maximal number of phonons we can encode in that bath. Therefore, it is natural to investigate the Hamiltonian for the special case $q=1$ and grow $k$ until convergence in the particle imbalance is reached. While this is a further simplification it is a good toy model for the system in question which, we expect, qualitatively inherits the basic properties of the effect in question. 
Moreover, while throughout this work we do not consider interactions of the spin system to environments other than the phonon degrees of freedom, 
the effect of the lab temperature, when sufficiently low, may to a crude approximation be invoked to argue for a cut of in the bosonic degrees of freedom.
We further set the disorder strength $W=10$ in order to find sufficiently good MBL as we restrict to small systems $n=8$. 

For weak couplings $\widetilde{g}\lesssim 0.3$ and arbitrary small oscillator energy $\widetilde{\omega}$ we find the spin localization to be almost unaffected by the presence of phonons 
(c.f.~Eq.~(\ref{fig:phonon-coupling}) (a) and (f)). 
Due to the small effective phonon energy $\widetilde\omega$ the number of phonons may grow to a non-negligible value. However, the coupling $\widetilde g$ is too weak compared to the robustness of the spin system in order leave a significant effect on the localization behaviour.
Increasing $\widetilde{g}\approx5$ we find that the localization of the spin degrees of freedom vanishes and that the phonon count grows as we increase $k$, even for small but finite $0<\widetilde{\omega}\lesssim10$: 
Evidence for the absence of MBL in this regime (c.f.~Eq.~(\ref{fig:phonon-coupling}) (b) and (g)). 
Increasing $\widetilde{\omega}\gtrsim10$ with $\widetilde g=5$ fixed we find that the spin degrees of freedom are brought back to localization. In particular, the phonon occupation gets suppressed and the imbalance is stabilized, as $\widetilde{\omega}\gtrsim50$ grows sufficiently large (c.f.~Eq.~(\ref{fig:phonon-coupling}) (c)-(e) and (h)-(j)).
This confirms our intuition from the previous section, that spin or electron localization can be made robust against phonons through a sufficiently high mobility edge. 

\section{Conclusion and outlook} 

In this work, we have studied a novel mechanism to design and synthesize real materials that are likely to display 
many body localized phenomena. Moreover, we have analyzed the effect of the electron-phonon coupling on the localization phenomenology, to provide evidence that properties of the MBL phase 
may be robust under experimentally realistic circumstances. 
We find substantial evidence that materials arising out of doping two Heisenberg magnets with a strong Ising anisotropy indeed
gives rise to many body localization. 

In particular, we find that a necessary ingredient is that the two magnets are at the same time sufficiently Ising anisotropic, and sufficiently different in terms of their Ising coupling. The difference of Ising couplings of the two materials together with the pronouncedness of the Ising anisotropy controls the quality of the MBL phenomenology. 
While the doping procedure can also lead to randomness in the hopping term of the resulting material, we find that randomness in this term does not play a significant role on the localizing property of the material, provided the hopping terms remain significantly smaller than the Ising terms of our  effective Hamiltonian. When the hopping terms dominate, our system tends  to delocalize.
The doping ratio on the other hand has a similar yet weaker effect on the quality of MBL phenomenology. Our numerical results indicate that the resulting material may not be many body localized for combinations with too weak doping ratios. 

As a practical example, we have analyzed the likely outcome of doping different real materials in the lab. The first is the compound  Sr$_{1-\delta}$Ba$_\delta$Co$_2$V$_2$O$_8$ made from the compounds SrCo$_2$V$_2$O$_8$ and BaCo$_2$V$_2$O$_8$. Our analysis predicts that the resulting compound is  most likely to thermalize. This is because the Ising anisotropy in the first compound is not sufficiently pronounced. 
In a second example, we considered doping between the materials CsCoCl$_3$ and CsCoBr$_3$ \cite{nagler_ising-like_1983}. The situation is more complicated here and the results may look localizing for finite system size. However, since the difference between the Ising coupling terms of the two materials does not seem to be sufficiently pronounced, the resulting material CsCoCl$_{3(1-\delta)}$Br$_{3\delta}$ is unlikely to display properties of MBL as evidenced in our numerical analysis.

We have further explored the effect of the electron-phonon coupling on the many body localized spin phenomenology. To this aim we derived a novel effective theory describing the spin phonon coupling making small scale numerical simulations of the global system more accessible. Our main theoretical insight is that the randomness in the spin Hamiltonian for the set-up we investigated will be reflected in a random distribution of local phonon energies and phonon-phonon couplings. Hence, for sufficiently strong disorders the phonon Hamiltonian will be described by an Anderson localized Hamiltonian. 

Using results for Anderson localization in truly one dimensional systems we argue that the many body localized phenomenology will persist after coupling the spin Hamiltonian to the corresponding phonon system for sufficiently strong disorder. 

Three dimensional systmes, where we think of wires of one dimensional spin chains  embedded in a three dimensional lattice, are particularly intricate. 
Due to the mobility edge of the three dimensional Anderson localized phonon system---which is controlled by the 
disorder 
strength---there will exist non-local phonon states. However, due to their large phonon energy their coupling with spin degrees of freedom will be effectively reduced. We derived an effective Hamiltonian that represents the spin Hamiltonian coupled to a high energy phonon and numerically confirmed that a large phonon energy can counter the effects of a large electron-phonon coupling. We, therefore, believe that for sufficiently strong disorder, artefacts of the many body localization will persist after coupling the spin system to the correspondingly disordered phonon system. In particular, the spin transport should be reduced depending on the disorder at sufficiently low energies.  This is phenomenologically in good agreement with the results of Ref.~\cite{nguyen_signature_2020}. 

A similar chain of reasoning works in two dimensions, being in good agreement with observations about thin film experiments made in
Ref.~\cite{ovadia_electron-phonon_2009,OvadiaSciRe2015}.

Our work can be understood in two ways. 
To start with, we provide a road map for experimentalists for creating materials that carry signs of many body localized phenomena. In particular we suggest concrete properties for materials such that the mix of these is believed to display many body localized phenomena. Our 
effective theory for the electron-phonon coupled system further gives a physical intuition that may lend itself towards the design of experimentally tractable measurement set-ups to confirm or falsify the claims made.

In a different reading, we re-analyze the effect of the electron-phonon coupling present in realistic materials on MBL. Contrary to the common belief, however, we arrive at the conclusion that meaningful artefacts of many body localization are actually possible in real materials. In this sense, our work can be understood as pointing out concrete properties and assumptions that remain to be analyzed in more detail in order to arrive at a final answer as to whether many body localization or artifacts thereof can survive when transiting from the theoretical description to an actual realization of such systems in real materials. 

Either way, our work opens the possibility and creates a clear map for future research to settle the question of whether real materials can exist carrying signs of many body localization in a more rigorous manner. It would also be interesting to consider situations where we mix more than two ingredient compounds giving rise to higher levels of discrete disorder. Another possibility is to consider spin chains of higher spins in order to increase the range of candidate materials that can give rise to real materials with properties of MBL. It is the hope that the present work can bring MBL in real quantum materials substantially closer to reality.

\section*{Acknowledgements}

We would like to acknowledge discussions with
Steven Thomson,
Christian Bertoni,
Marcel Goihl,
Curt von Keyserlingk,
Christian Krumnow and
Bohan Lu. 
A.~N.~is 
very thankful to the local IT team and in particular to J\"org Behrmann and Jens Dreger for generous 
support regarding the numerical implementation and maintenance of the computational resources. 
Our tensor network simulations were performed using the TeNPy library~\cite{tenpy}.
We thank the DFG (CRC 183 and FOR 2724) 
for funding. This work also touches upon the themes of 
BMBF (MUNIQC-Atoms) in the tensor network method development.

\appendix

\begin{figure*}
    \centering
    \includegraphics[width=.9\linewidth]{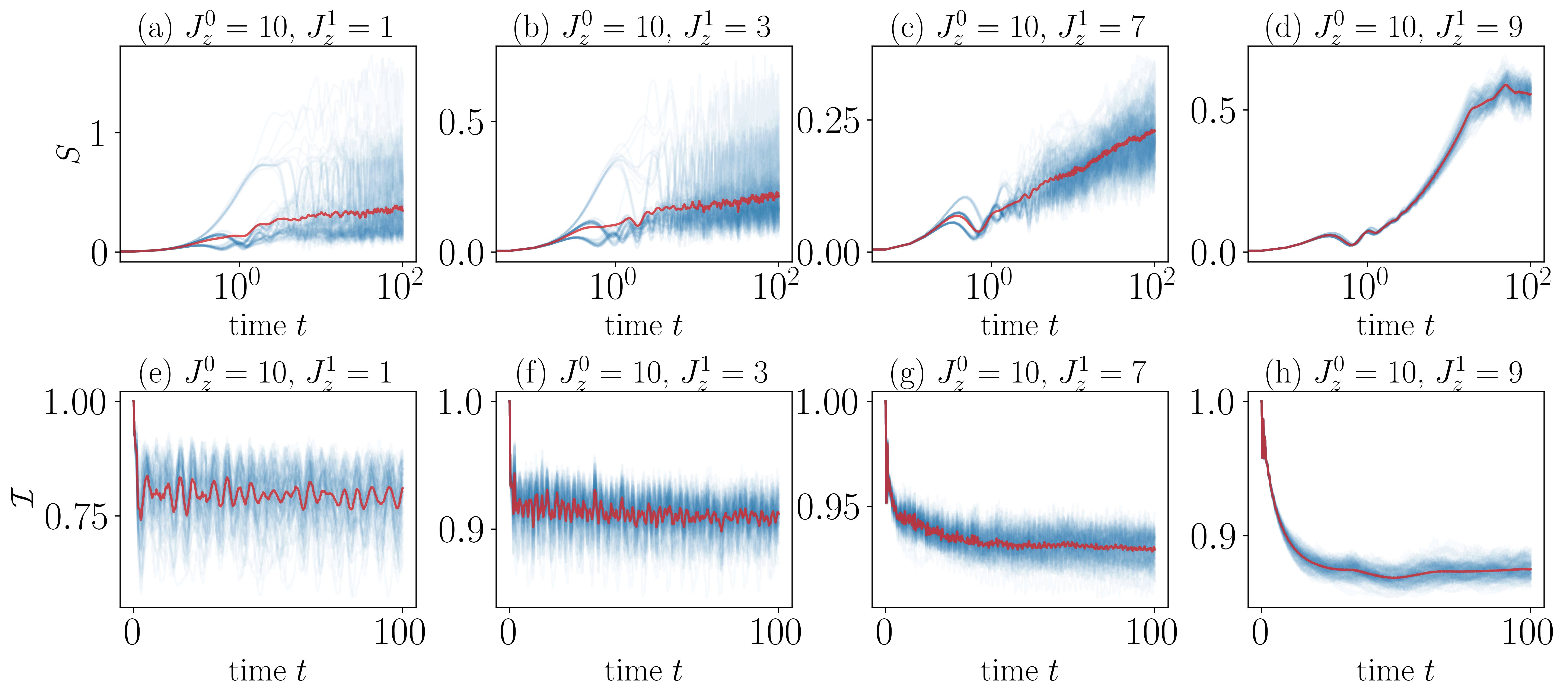}
    \caption{Entanglement entropy (top row) and particle imbalance (bottom row) for different interaction gaps $J_z^0-J_z^1$ at doping strength $\delta=0.3$ and $J_\perp^i=1$. In all plots $J_z^0=10$ is fixed and $J_z^1$ is varied as
    $J_z^1=1$ in (a), (e), 
    $J_z^1=3$ in (b), (f), 
    $J_z^1=7$ in (c), (g), 
    and
    $J_z^1=9$ in (d), (h).}
    \label{fig:gap_scaling}
\end{figure*}

\section{Supplemental material to the numerical investigation}

\subsection{On the role of the interaction strength gap}
\label{app:interaction-gap}
In this appendix, we investigate the role of the interaction strength gap $\Delta=J_z^0-J_z^1$. Figure \ref{fig:gap_scaling} shows the data for fixed doping strength $\delta=0.3$, $J_\perp^i=1$ and $J_z^0=10$ for varying $J_z^1=1,3,7,9$. The variance of the individual plots becomes vanishing as $\Delta$ becomes small as in 
Fig.~\ref{fig:gap_scaling} (d) and (h) ($J_z^1=0.9$). This indicates that the corresponding system behaves approximately like the homogeneous, undoped, material as analyzed in 
Appendix
\ref{app:finite-size-scaling}.

Figures \ref{fig:gap_scaling} (b), (c), (f) and (g) resemble the corresponding plots for $J_z^1=5$ as in the main text and can be seen as evidence for MBL.
As $J_z^1$ becomes as small as $1$ in 
Figs.~\ref{fig:gap_scaling} (a) and (e) the effect becomes less pronounced. We expect that this is due to the proximity of the hopping strength and the interaction strength of the second material.\\

\subsection{Finite size scaling for strong Ising anisotropy}
\label{app:finite-size-scaling}

\begin{figure}
    \centering
    \includegraphics[width=.8\linewidth]{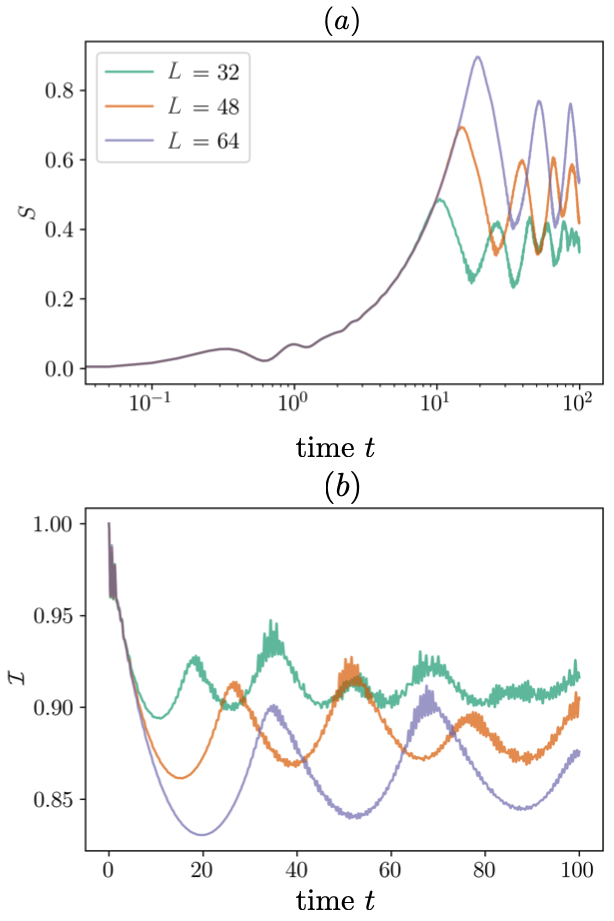}
    \caption{Finite size scaling for the entanglement entropy (a) and the imbalance (b) in the  homogeneous strong Ising anisotropic XXZ Hamiltonian.}
    \label{fig:ai_scaling}
\end{figure}

In this appendix, we analyze the homogenous, i.e., undoped XXZ Hamiltonian at strong Ising anisotropy. As shown in Fig.~\ref{fig:ai_scaling} we find that the the entanglement entropy as well as the particle imbalance  saturates at any fixed system size to a finite value. This value, however, increases, respectively decreases, as the system size is increased. In particular, we do not expect to find a saturating particle imbalance in the thermodynamic limit. This agrees with the fact that the homogenous XXZ model at any finite anisotropy is ergodic.

\bibliographystyle{apsrev}

\end{document}